# Research on College Students' Innovation and Entrepreneurship Education from The Perspective of Artificial Intelligence Knowledge-Based Crowdsourcing

Yufei Xie, Xiang Liu, Qizhong Yuan
Jiangxi College of Applied Technology, Ganzhou, Jiangxi 341000, China

**Abstract:** Based on the practical process of innovation and entrepreneurship education for college students in the author's university, this study analyzes and deconstructs the key concepts of AI knowledge-based crowdsourcing on the basis of literature research, and analyzes the objective fitting needs of combining AI knowledge-based crowdsourcing with college students' innovation and entrepreneurship education practice through a survey and research of a random sample of college students, and verifies that college students' knowledge and application of AI knowledge-based crowdsourcing in the learning and practice of innovation and entrepreneurship The study also verifies the awareness and application of AI knowledge-based crowdsourcing knowledge by university students in the learning and practice of innovation and entrepreneurship.

## 1. INTRODUCTION

With the advancement of globalization and the widespread use of information technology, crowdsourcing has gradually become a major way for society at large to participate in Internet activities. howe (2006) was the first to introduce the concept of crowdsourcing, which is considered to be the practice of a company or organization to outsource, on a free and voluntary basis, work tasks that used to be performed by employees to a non-specific (and often large) mass networks to find solutions.

In recent years, artificial intelligence technology has developed at a high speed, and big data processing increasingly requires high accuracy and coverage. Human-machine hybrid intelligence is still the main form of current artificial intelligence development. In the process of developing AI from perception to cognition, a large number of human interventions are required in the creation, acquisition, annotation, refinement and other work aspects of knowledge. The requirements of these work segments are no longer met by the traditional labour-based crowdsourcing (labour-intensive crowdsourcing) model, with high business hierarchy, high intellectual threshold and high quality requirements, which has led to the development of knowledge-based crowdsourcing.

The development needs of the knowledge-based crowdsourcing model fit well with the innovation and entrepreneurship education needs of university students, such as comprehensive literacy, innovative thinking and entrepreneurial practice. In the business processes and scenarios of knowledge-based crowdsourcing, it is very easy for university students to find the role of the group and adapt to the business model of knowledge-based crowdsourcing relatively quickly, opening up a new direction of research and exploration for innovation and entrepreneurship education for university students in the new era.

## 2. OVERVIEW OF KNOWLEDGE-BASED CROWDSOURCING

Crowd-sourcing (Crowd-sourcing) is a new outsourcing model that links a loose group of task issuers (Requester) and task completers (workers, Worker) to achieve a series of operations such as task issuance, matching, completion and payment. [1] Traditional labour-based crowdsourcing, characterised by a single task, simple task evaluation methods, single worker requirements, low worker threshold and relatively large number of workers, mainly organises idle labour to serve the public, with specific task scenarios such as delivery of courier, drop-off takeaway, distribution of leaflets and handcrafting.

Knowledge-based crowdsourcing, which emerged with the development of artificial intelligence technology, is a concrete manifestation of human-computer hybrid intelligence, which is a rational organization of the public's wisdom and time to provide human intellectual support for projects, and is characterized by strong task diversity, worker diversity, difficulty in evaluating task quality, and a large impact surface of task completion quality, etc. The main task scenarios are image annotation, language recognition, general knowledge collection, question answering, etc. The incentive mechanism of labour-based crowdsourcing mainly relies on material rewards such as cash and tokens, while the incentive mechanism of knowledge-based crowdsourcing is more about spiritual rewards such as self-worth, self-improvement and self-pleasure, rather than just material rewards. [2]

## 3. THE INTEGRATION OF INNOVATION AND ENTREPRENEURSHIP EDUCATION WITH KNOWLEDGE-BASED CROWDSOURCING

With the rapid development of artificial intelligence, "Internet+", big data and other technologies, the gradual implementation of the innovation-driven national strategy has vigorously promoted the rapid development of innovation and entrepreneurship education in China. [3] The innovation and entrepreneurship education of college students has gradually developed from the early theoretical teaching-based innovation and entrepreneurship knowledge education, such as financial taxation, laws and regulations, successful cases and sandbox simulation, to the application-oriented comprehensive literacy education based on knowledge education, using information technology as a means and project practice as a grip.

The main problems of innovation and entrepreneurship education at present are, firstly, insufficient publicity on innovation and entrepreneurship education, and the actual number of university students who can participate in the whole process is still a minority; secondly, the investment in innovation and entrepreneurship education needs to be enhanced, and it is difficult to solve the investment problem by relying on universities themselves alone; thirdly, the form of innovation and entrepreneurship education is old-fashioned and single, and practicality and diversity are still insufficient.[4]

From the perspective of the characteristics of knowledge-based crowdsourcing, on the one hand, with the absolute number of university students in China, there is a high probability that outstanding individuals will emerge as task issuers of knowledge-based crowdsourcing; on the other hand, in terms of university students as a group, they have characteristics such as higher intellectual level, higher willingness to upgrade, lower material needs, better organisation initiation and more flexible time arrangement, etc. These characteristics of university students are highly compatible with the labour-based crowdsourcing These characteristics of university students are highly compatible with the characteristics of labour-based crowdsourcing, and the role of university students and task finishers are a perfect match.

In the process of development, knowledge-based crowdsourcing can be used as the basics of innovation and entrepreneurship education for university students, as a practical mode of innovation and entrepreneurship education for university students, and as a new channel for university students to work hard and gain income. By linking up with knowledge-based crowdsourcing projects and platforms, and educating and guiding university students to participate in knowledge-based crowdsourcing business practices, universities can effectively solve the problems encountered in the current innovation and entrepreneurship education.

## 4. UNIVERSITY STUDENTS' PERCEPTIONS OF KNOWLEDGE-BASED CROWDSOURCING

The author conducted a knowledge-based crowdsourcing awareness questionnaire with 835 students enrolled in a university and the results were analysed as follows.

Table 1 Results of the question Do you know about knowledge-based crowdsourcing as an entrepreneurial approach?

| options | subtotal | The proportion |
|---|---|---|
| Very familiar with | 43 | 5.15% |
| To understand | 140 | 16.77% |
| Don't understand | 652 | 78.08% |

Many university students do not know about knowledge-based crowdsourcing, and those who do know about it have only heard about it in a general way. Promoting innovation and entrepreneurship among university students through knowledge-based crowdsourcing has not yet been officially carried out in universities.

Table 2 Results of the survey on what you would like to find out about knowledge-based crowdsourcing

| options | subtotal | The proportion |
|---|---|---|
| Willing to | 660 | 79.04% |
| Don't want to | 175 | 20.96% |

Nearly 80% of university students are curious about knowledge-based crowdsourcing, and they want more channels and ways for them to learn about and acquire the knowledge and skills.

Table 3 Results of the question on what kind of knowledge-based crowdsourcing projects you are interested in

| options | subtotal | The proportion |
|---|---|---|
| Image annotation | 147 | 17.6% |
| Language identification | 102 | 12.2 2% |
| Common sense collection | 149 | 17.84% |
| Are willing to contact | 437 | 52.3 4% |

The fact that university students are more interested in the main task forms of knowledge-based crowdsourcing, with more than half of them choosing to be willing to engage with them, also proves that knowledge-based crowdsourcing is converging with the values of university students.

## 5. EXPLORING THE PRACTICE OF KNOWLEDGE-BASED CROWDSOURCING INNOVATION AND ENTREPRENEURSHIP EDUCATION

In the practice of innovation and entrepreneurship education in the second half of 2021, the author introduced knowledge-based crowdsourcing projects such as map marking and image recognition through a third-party organization and organized some university students to participate in the practice. Through the practice and the feedback from the participating university students, the objective fitting needs of combining knowledge-based crowdsourcing with the innovation and entrepreneurship education practice of university students were analysed comprehensively as.

### 5.1 Universities need to truly recognise the role of knowledge-based crowdsourcing in innovation and entrepreneurship education

Through research and practice, I found that the biggest problem that hinders university students from participating in knowledge-based crowdsourcing innovation and entrepreneurship education is that they do not understand knowledge-based crowdsourcing itself, neither the basic knowledge of knowledge-based crowdsourcing, nor the practical way of knowledge-based crowdsourcing projects. The fact that most students do not understand it then also means that universities themselves do not understand and pay attention to knowledge-based crowdsourcing, and do not keep up with the times by closely combining the direction of advanced productivity development, such as artificial intelligence, with school education and teaching. Only when universities realize that knowledge-based crowdsourcing is an important way and method to solve the current innovation and entrepreneurship education in the field of innovation and entrepreneurship education, can the university student population be fully exposed to the knowledge and projects of knowledge-based crowdsourcing, and the human resource base of knowledge-based crowdsourcing can be formed quickly.

5.2 University students need to be objective about knowledge-based crowdsourcing innovation and entrepreneurship practices

In the practice of knowledge-based crowdsourcing innovation and entrepreneurship, I found that the knowledge and skill levels of university students involved in knowledge-based crowdsourcing projects are fully competent, and many of them are reflected in practice. However, there are still many university students who stop halfway and do not complete the set task objectives. There are three main reasons for this analysis: firstly, the participating university students subjectively have insufficient knowledge of knowledge-based crowdsourcing projects and have participated in projects they are not good at or are not psychologically prepared for long-term practice; secondly, a few university students' perceptions of the benefits of innovation and entrepreneurship practice deviate from the actual situation and have not yet been able to understand the hardship of entrepreneurial work; thirdly, this practice is experimental in nature and the practice projects that can be chosen for university students are The third is that the practice is experimental in nature and the choice of practical projects for students is very limited.

5.3 Crowdsourcing platforms need to efficiently aggregate information and resources for crowdsourcing tasks

The practice is a knowledge-based crowdsourcing business that the author contacted a third-party agency to match, and there is currently no information platform on the internet that can consistently provide knowledge-based crowdsourcing. From another perspective, knowledge-based crowdsourcing is still in its infancy in the whole market, especially for university students. Although they value spiritual incentives, it does not mean that they do not need material incentives. In the initial construction of knowledge-based crowdsourcing platforms such as Encyclopedia and Zhihu, netizens are still mainly involved in problem solving for free. However, in order to truly implement the knowledge-based crowdsourcing model, more crowdsourcing businesses with material rewards must be able to be aggregated and distributed through the online platform, so that they can be efficiently matched with the university student population from the model.

6. CONCLUSION

Knowledge is the fruit of human understanding of the world and its transformation. No matter how the times change and how technology advances, advanced technological means exist for the progress of human civilization and the growth of human beings themselves. In the context of AI knowledge-based crowdsourcing, there are still very few research results on innovation and entrepreneurship education for university students, and there is a wide range of areas that can be researched, and there is also a lot of work that can be done to guide university students to carry out knowledge-based crowdsourcing innovation and entrepreneurship practice in parallel with theoretical research. Knowledge-based crowdsourcing provides a new practical grasp for the reform and development of university students' innovation and entrepreneurship education, and opens up a new exploration path for comprehensively improving university students' innovation and entrepreneurship ability in the context of the era of great development of artificial intelligence.

ACKNOWLEDGEMENTS
The 2021 Ganzhou City Social Science Research Project (2021-028-0391).